\documentclass[prb,twocolumn,superscriptaddress]{revtex4}

\usepackage{graphicx}
\usepackage{bm}
\usepackage{amsmath}
\usepackage{amssymb}
\usepackage{hyperref}

\begin{document}

\title{Diameter dependence of SiGe nanowire thermal conductivity}

\author{Zhao Wang}
\email{wzzhao@yahoo.fr}
\affiliation{LITEN, CEA-Grenoble, 17 rue des Martyrs, 38054 Grenoble Cedex 9, France}

\author{Natalio Mingo}
\affiliation{LITEN, CEA-Grenoble, 17 rue des Martyrs, 38054 Grenoble Cedex 9, France}

\begin{abstract}
We theoretically compute the thermal conductivity of SiGe alloy nanowires as a function of nanowire diameter, alloy concentration, and temperature, obtaining a satisfactory quantitative agreement with experimental results. Our results account for the weaker diameter dependence of the thermal conductivity recently observed in Si$_{1-x}$Ge$_x$ nanowires ($x<0.1$), as compared to pure Si nanowires. We also present calculations in the full range of alloy concentrations, $0 \leq x \leq 1$, which may serve as a basis for comparison with future experiments on high alloy concentration nanowires.

\end{abstract}

\maketitle

The potential interest of nanowires as thermoelectric materials has been manifest for more than a decade.\cite{Cahill2003,Hicks1993,Heremans2000} Although interest was initially motivated by hopes of taking advantage of electron confinement in the structures, it soon became clear that another advantage of nanowires was their potentially strongly reduced thermal conductivity. Advanced techniques enabled the measurement of thermal conductivity of single nanowires.\cite{Shi2003} Large reductions in Si nanowire lattice thermal conductivity were experimentally reported in 2003, further stimulating research activities in this area.\cite{Li2003} Some astonishingly low thermal conductivities have also been recently claimed on Si nanowires.\cite{Hochbaum2008}

Many of the initial investigations on thermal conductivity reduction by nanostructuring had concentrated on ordered crystalline structures. However in recent years, the advantages of nanostructured alloys have been underlined for various systems, such as embedded nanodots,\cite{Kim2006,Mingo2009} or nanoporous materials.\cite{Bera2010} These works show that the interplay between alloy scattering and scattering by the nanostructured features can lead to interesting qualitative differences between the behavior of the thermal conductivity $\kappa$ of alloy and non-alloy structures. In particular, a slower dependence of $\kappa$ on nano-featured size is expected when using alloys. Similar effects may thus take place in SiGe nanowires, due to the interplay between alloy and boundary scattering. Although the thermal conductivity of SiGe nanowires was measured in Ref.\onlinecite{Li2003a}, no important reduction was reported there, presumably due to the low total Ge concentration. It is only very recently that the thermal conductivity of homogeneous SiGe nanowires with $0<x<0.1$ have been reported, obtaining remarkable reductions in $\kappa$ below the bulk alloy value.\cite{Kim2010}

In addition to the small reported $\kappa$ values, the experimental study stressed their much weaker diameter dependence as compared with pure Si nanowires. Thus, the question is whether one can theoretically describe and quantify these effects in agreement with the experimental results. In this Letter, we address this problem, and we provide theoretical curves for the dependence of $\kappa$ on nanowire diameter, alloy concentration, and temperature. In addition, we also calculate results for Ge concentrations higher than the ones experimentally measured so far, allowing for future experimental comparison with Ge rich samples.

A full dispersion theoretical model for the thermal conductivity of Si nanowires was presented in Ref.\onlinecite{Mingo2003}. Such a model only required prior knowledge of bulk Si experimental thermal conductivity, but did not rely on fitting to nanowire measurements. Computing the full phonon dispersions may be laborious. However, as it was noted in Ref.\onlinecite{Mingo2003}, a very good description of nanowire thermal conductivity can be obtained without the need to compute the full dispersions, just by introducing an additional adjustable parameter, the cutoff frequency $\omega_{c}$, combined with a much simpler linear dispersion approximation. 

In this simpler approach, the cutoff frequency is the only parameter that relies on nanowire measurements. But to fit it, it usually suffices to have measurements of $\kappa(T)$ for just one single nanowire, and one is then able to calculate the thermal conductivity of nanowires of other diameters without any additional fitting. The simplicity of this method makes it very attractive for the analysis of other systems. In particular, it can be employed to predict the thermal conductivity of SiGe nanowires \textit{without any prior knowledge of the experimental measurements}. To do so, only two more things are needed: knowledge of the experimental $\kappa(T)$ for bulk Ge, and knowledge of $\kappa(x)$ for Si$_{1-x}$Ge$_x$ at a given temperature.

Basis of this calculation approach are provided in Refs.\onlinecite{Mingo2003} and \onlinecite{Mingo2009}, a summary is given here. The thermal conductivity is computed as

\begin{eqnarray}
\label{eq:6}
\kappa =  \frac{k_{B}^{4} T^{3}}{2\pi^{2} v_{b} \hbar^{3}} \int_{0}^{\frac{\hbar\omega_{c}}{k_{B} T}}{\tau(T,y) y^4 {\frac{e^{y} }{(e^{y} -1)^{2}}}dy}.
\label{eq:cond}
\end{eqnarray}

Denoting the Si concentration by $x$, the different magnitudes in the above equation are:
$v_{b}^{-2} = xv_{b,Si}^{-2} + (1-x)v_{b,Ge}^{-2}$, $\omega_{c} = \omega_{c,Si}(v_{b}/v_{b,Si})$, $k_{B}=$ Boltzmann's constant, $\hbar=$ reduced Planck's constant, $T=$ temperature, and $y\equiv \hbar\omega/ k_{B} T$. The average speeds of sound for Si and Ge are obtained from the experimental sound velocities of the transverse and longitudinal acoustic branches, $c_T$ and $c_L$, as $v_{b,Si(Ge)}^{-2} = (2/3)(c_{T,Si(Ge)})^{-2}+(1/3)(c_{L,Si(Ge)})^{-2}$. The cutoff frequency for Si was adjusted in Ref.\onlinecite{Mingo2003}, to be about 40 THz. The scattering rate $\tau(\omega)$ is expressed using Mathiessen's rule, as a combination of anharmonic ($\tau_u$), alloy ($\tau_a$), and boundary ($\tau_b$) scattering contributions:

\begin{eqnarray}
\tau^{-1} = \tau_{u}^{-1} + \tau_{a}^{-1} + \tau_{b}^{-1}.
\end{eqnarray}

Following the virtual crystal approximation, the anharmonic contribution is approximated as a linear interpolation between Si and Ge:

\begin{eqnarray}
\tau_u^{-1} = x\tau_{u,Si}^{-1} + (1-x)\tau_{u,Ge}^{-1},
\end{eqnarray}

with $\tau^{-1}_{u,Si} = B_{Si}\omega^2 T e^{C_{Si}/T}$,\cite{Mingo2003} and similarly for Ge. Parameters $B$ and $C$ were adjusted to fit the bulk experimental $\kappa(T)$ curves\cite{Steele1958} (see Table \ref{table1}).

\begin{figure}[ht]
\centerline{\includegraphics[width=9cm]{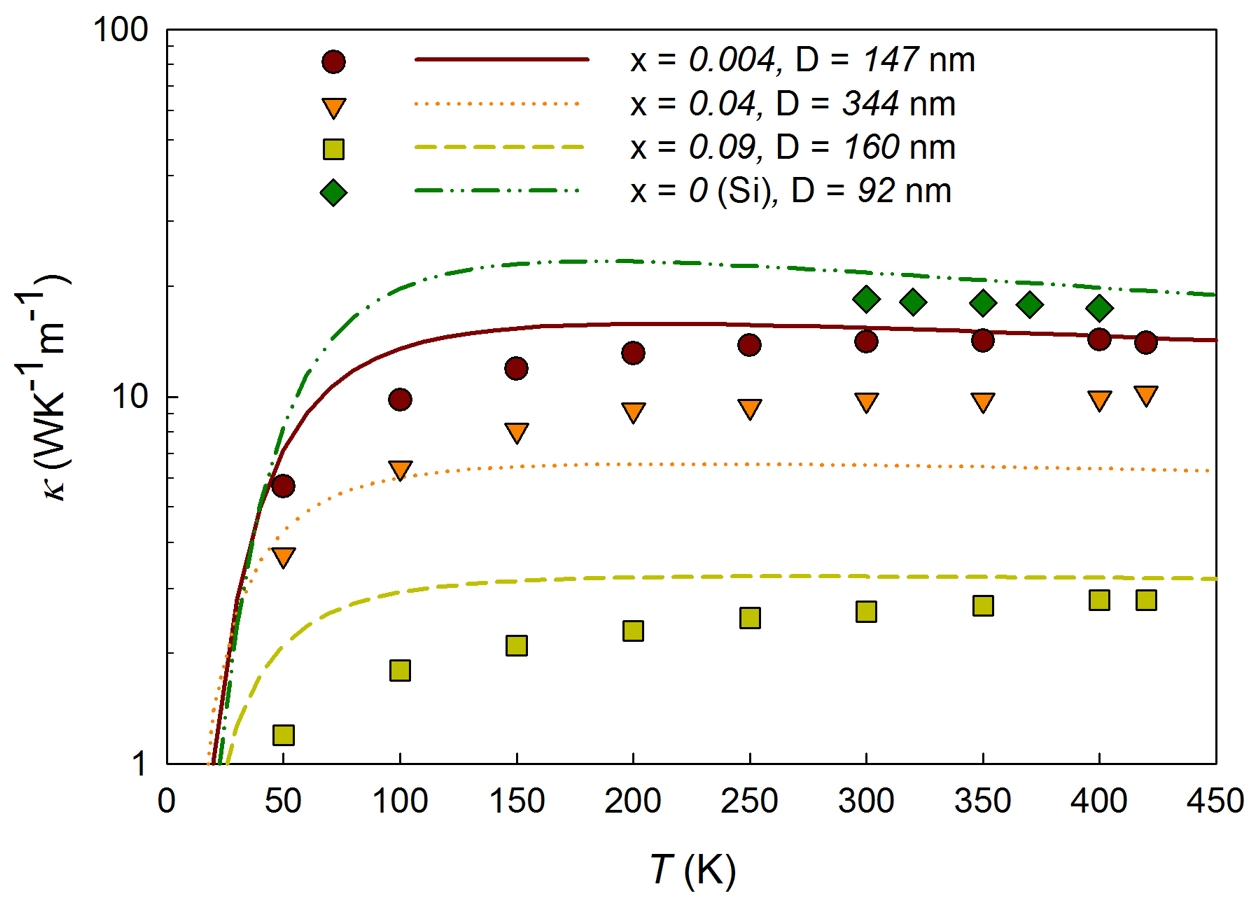}}
\caption{\label{fig:T1}
(Color online) Thermal conductivity $\kappa$ \textit{vs.} temperature $T$ for Si$_{1-x}$Ge$_{x}$ NWs. $x$ is the germanium concentration and $D$ is the NW diameter. The lines show calculation results using Eq.\ref{eq:6}, the symbols represent experimental results from Ref.\onlinecite{Kim2010}.}
\end{figure}

The alloy scattering term is derived as

\begin{eqnarray}
\tau_a^{-1} = x(1-x)A\omega^4,
\end{eqnarray}

where constant $A$ was adjusted to measurements of $\kappa(x)$ of bulk SiGe alloys for giving the best fitting to experimental data.\cite{Steele1958}  Boundary scattering is included as $\tau_b^{-1} = v_{b}/D$, where $D$ is the nanowire diameter. The model is thus the same as in Ref.\onlinecite{Mingo2009}, with the difference that the nanoparticle scattering rate in that reference is here substituted by the boundary scattering term. Here we use the same parameters as in that reference, except for $A$, which is now fitted by the values of Ref.~\onlinecite{Steele1958}, rather than those of Abeles.\cite{Abeles1962} Thus, no attempt has been made to fit the nanowire measurements from Ref.\onlinecite{Kim2010} via adjustable parameters. The very reasonable agreement with those results (see below) is quite remarkable, given the simplicity of the model and the various approximations involved.

\begin{table}[h]
\caption{\label{table1} Table of parameters.}
\begin{center}
\begin{tabular}{lr}
   parameter &   value (unit) \\
\hline 
   $v_{b,Si}$           &   $6400$ (m/s) \\
   $v_{b,Ge}$           &   $3900$ (m/s) \\
   $\omega_{c,Si}$      &   $38.8$ (THz) \\
   $A$                  &   $3.01 \times 10^{-41}$ (s$^{3}$) \\
   $B_{Si}$             &   $1.51 \times 10^{-19}$ (s/K) \\
   $B_{Ge}$             &   $2.91 \times 10^{-19}$ (s/K) \\
   $C_{Si}$             &   $139.8$ (K) \\
   $C_{Ge}$             &   $69.34$ (K) \\
\end{tabular}
\end{center}
\end{table}

\begin{figure}[ht]
\centerline{\includegraphics[width=9cm]{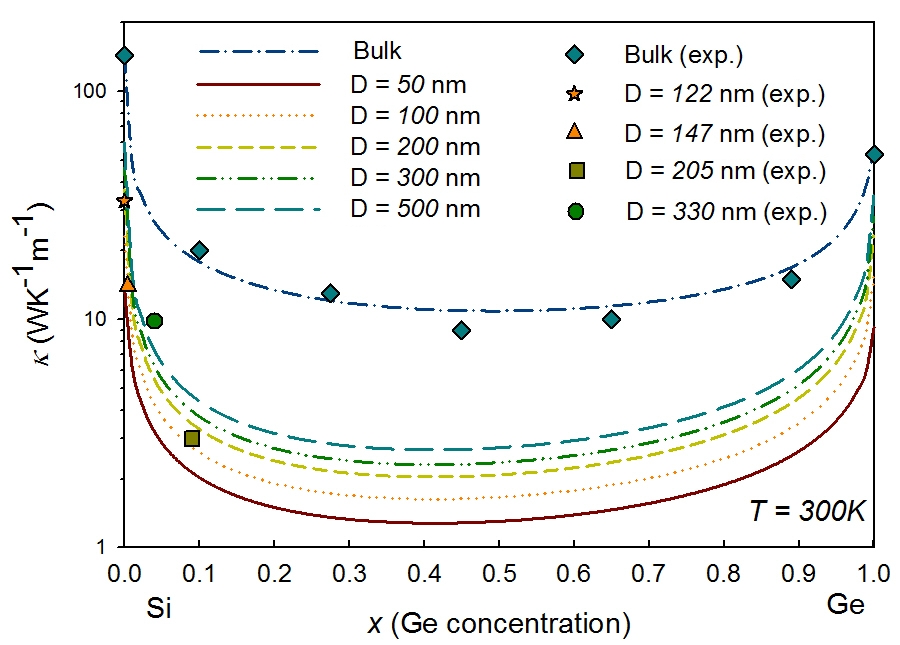}}
\caption{\label{fig:x}
(Color online) Thermal conductivity \textit{vs.} germanium concentration for Si$_{1-x}$Ge$_{x}$ bulk alloy and NWs with different diameters. The lines show calculation results using Eq.\ref{eq:6}, and the symbols represent experimental data from Ref.\onlinecite{Steele1958} and \onlinecite{Kim2010}.}
\end{figure}

A direct comparison between experimental data and our calculated results for four NWs is shown in Fig.\ref{fig:T1}.
Rather good quantitative agreement is obtained for the three cases with Ge concentration $x=0, 0.004$ and $0.09$, especially at temperatures above 200K. For the NW with $x=0.04$ and $D=344$nm, the difference between the theoretical curve and the experimental data reaches $0.3$ at room temperature. This may be due to the fact that $\kappa$ is very sensitive to $x$ when $x$ is small ($<0.05$, see  Fig.\ref{fig:x} and discussion below), i.e., a small deviation of $x$ can make important difference in $\kappa$.   

\begin{figure}[ht]
\centerline{\includegraphics[width=9cm]{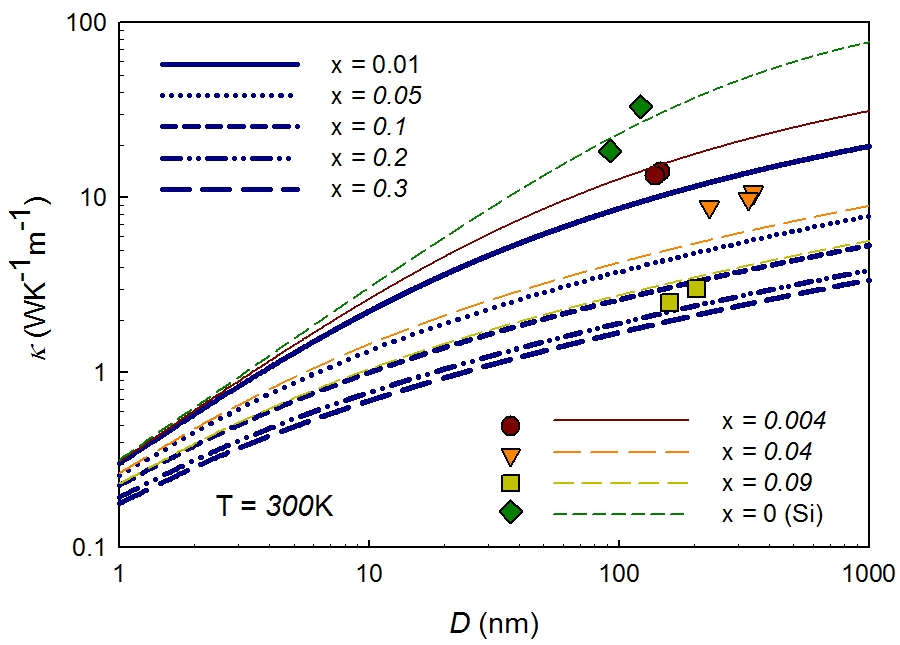}}
\caption{\label{fig:D}
(Color online) Thermal conductivity \textit{vs.} NW diameter (log. scale) for Si$_{1-x}$Ge$_{x}$ NWs with different germanium concentrations at room temperature. The lines show calculation results using Eq.\ref{eq:6}, the symbols represent experimental results from Ref.\onlinecite{Kim2010}.}
\end{figure}

We show the dependence of $\kappa$ on nanowire diameter in Fig.\ref{fig:D}. For small $D$, $\kappa$ is proportional to $D$, whereas this dependence becomes slower as $D$ increases. The deviation from linear dependence occurs at smaller $D$ the larger the Ge concentration, up to about $x\sim 0.5$. Obviously, as $x$ increases further, the situation reverts, with pure Ge nanowires displaying a large linearity range (not shown) similarly to pure Si nanowires. This effect is due to the coexistence of alloy and boundary scattering contributions, and it is totally analogous to the effect predicted on nanoporous materials in Ref.~\onlinecite{Bera2010}, where the role of thickness was played by the distance between pores. The slower dependence is related to the very fast frequency dependence of alloy scattering. Alloy scattering blocks high frequency phonons very effectively, but it is totally transparent to low frequency phonons. Thus, the thermal conductivity of an alloy is dominated by low frequency phonons with very long mean free paths, whereas in non-alloys $\kappa$ contains contributions from a larger range of frequencies with shorter mean free paths on average. 

Introducing a boundary therefore affects the thermal conductivity of an alloy already at rather large values of $D$, whereas for that same $D$ there is little effect on $\kappa$ of a non-alloy. In the limit of very small $D$, however, boundary scattering dominates over alloy and anharmonic scattering at all frequencies. In that regime, $\kappa$ of Si and SiGe become similar, because the effect of 'bowing' introduced by alloy scattering disappears. The competition between boundary and alloy scattering can be understood in analytical terms from Eq.~(\ref{eq:cond}). For high enough $T$ one can approximate $e^{y} \simeq 1+y$, so $\kappa\propto \int_0^{\omega_c}{\Big(v_{b}/D + A^{\prime} \omega^{4} \Big )^{-1}\omega^2d\omega}.$ with $A^{\prime}=x(1-x)A$.

For $D<v_{b} A^{^\prime-1}\omega_c^{-4}$, boundary scattering dominates at all phonon frequencies, and $\kappa\propto D$. When $D$ is larger than this, however, an upper range of frequencies becomes dominated by alloy scattering. The integral can be performed analytically, and it asymptotically tends to $\kappa\propto D^{1/4}$ for $D>>v_{b} A^{^\prime-1}\omega_{c}^{-4}$. Thus, the onset of the slower $D$ dependence of $\kappa$ is directly related to the Ge concentration: $D_{onset}\sim (x(1-x))^{-1}v_{b} A^{-1}\omega_c^{-4}$.
The aformentioned $D^{1/4}$  never clearly settles, since it is eventually superseded by anharmonic scattering for larger $D$, leading to the saturation of $\kappa$ towards its bulk value.

Finally we plot $\kappa$ as a function of $x$ in Fig.\ref{fig:x}. It is found that increasing the Ge concentration leads to a very fast decrease of thermal conductivity when $x<0.005$. Keeping increasing $x$, the curves of $\kappa$ then tend to saturate at a minimum value $\kappa_{min}$ around $x=0.4$. Crossing throught this minimum points, the thermal conductivities increase progressively with the increasing Ge concentration. This concentration dependence of NWs is very similar to the experimentally-observed one of bulk alloys.\cite{Steele1958} We note that $\kappa$ at $x=0.2$ is already close to $\kappa_{min}$. Moreover, it can beseen that $\kappa_{min}$ of bulk material is at least $7$ times larger than those of of NWs. 

In conclusion, we have presented the theoretical dependence of SiGe alloy nanowire thermal conductivity as a function of diameter, temperature, and Ge fraction. We establish the appearance of a slow diameter dependence regime beyond a certain onset diameter, which depends on the alloy concentration. The results explain the weak diameter dependence reported in a recent experiment, and are in reasonably good quantitative agreement with those experimental results. Results for higher Ge concentrations beyond the experimentally reported range have also been provided, and may allow for further testing of the theory by future experiments.


\begin{thebibliography}{14}
\expandafter\ifx\csname natexlab\endcsname\relax\def\natexlab#1{#1}\fi
\expandafter\ifx\csname bibnamefont\endcsname\relax
  \def\bibnamefont#1{#1}\fi
\expandafter\ifx\csname bibfnamefont\endcsname\relax
  \def\bibfnamefont#1{#1}\fi
\expandafter\ifx\csname citenamefont\endcsname\relax
  \def\citenamefont#1{#1}\fi
\expandafter\ifx\csname url\endcsname\relax
  \def\url#1{\texttt{#1}}\fi
\expandafter\ifx\csname urlprefix\endcsname\relax\def\urlprefix{URL }\fi
\providecommand{\bibinfo}[2]{#2}
\providecommand{\eprint}[2][]{\url{#2}}

\bibitem[{\citenamefont{Cahill et~al.}(2003)\citenamefont{Cahill, Ford,
  Goodson, Mahan, Majumdar, Maris, Merlin, and Phillpot}}]{Cahill2003}
\bibinfo{author}{\bibfnamefont{D. }~\bibnamefont{Cahill et al.}},
  \bibinfo{journal}{J. Appl. Phys.} \textbf{\bibinfo{volume}{93}},
  \bibinfo{pages}{793} (\bibinfo{year}{2003}).

\bibitem[{\citenamefont{Hicks and Dresselhaus}(1993)}]{Hicks1993}
\bibinfo{author}{\bibfnamefont{L.}~\bibnamefont{Hicks}} \bibnamefont{and}
  \bibinfo{author}{\bibfnamefont{M.}~\bibnamefont{Dresselhaus}},
  \bibinfo{journal}{Phys.\ Rev.~B} \textbf{\bibinfo{volume}{47}},
  \bibinfo{pages}{16631} (\bibinfo{year}{1993}).

\bibitem[{\citenamefont{Heremans et~al.}(2000)\citenamefont{Heremans, Thrush,
  Lin, Cronin, Zhang, Dresselhaus, and Mansfield}}]{Heremans2000}
\bibinfo{author}{\bibfnamefont{J.}~\bibnamefont{Heremans}},
  \bibinfo{author}{\bibfnamefont{C.}~\bibnamefont{Thrush}},
  \bibinfo{author}{\bibfnamefont{Y.-M.} \bibnamefont{Lin}},
  \bibinfo{author}{\bibfnamefont{S.}~\bibnamefont{Cronin}},
  \bibinfo{author}{\bibfnamefont{Z.}~\bibnamefont{Zhang}},
  \bibinfo{author}{\bibfnamefont{M.}~\bibnamefont{Dresselhaus}},
  \bibnamefont{and}
  \bibinfo{author}{\bibfnamefont{J.}~\bibnamefont{Mansfield}},
  \bibinfo{journal}{Phys.\ Rev.~B} \textbf{\bibinfo{volume}{61}},
  \bibinfo{pages}{2921} (\bibinfo{year}{2000}).

\bibitem[{\citenamefont{Shi et~al.}(2003)\citenamefont{Shi, Li, Yu, Jang, Kim,
  Yao, Kim, and Majumdar}}]{Shi2003}
\bibinfo{author}{\bibfnamefont{L.}~\bibnamefont{Shi}},
  \bibinfo{author}{\bibfnamefont{D.}~\bibnamefont{Li}},
  \bibinfo{author}{\bibfnamefont{C.}~\bibnamefont{Yu}},
  \bibinfo{author}{\bibfnamefont{W.}~\bibnamefont{Jang}},
  \bibinfo{author}{\bibfnamefont{D.}~\bibnamefont{Kim}},
  \bibinfo{author}{\bibfnamefont{Z.}~\bibnamefont{Yao}},
  \bibinfo{author}{\bibfnamefont{P.}~\bibnamefont{Kim}}, \bibnamefont{and}
  \bibinfo{author}{\bibfnamefont{A.}~\bibnamefont{Majumdar}},
  \bibinfo{journal}{J. Heat Transfer} \textbf{\bibinfo{volume}{125}},
  \bibinfo{pages}{881} (\bibinfo{year}{2003}).

\bibitem[{\citenamefont{Li et~al.}(2003{\natexlab{a}})\citenamefont{Li, Wu,
  Kim, Shi, Yang, and Majumdar}}]{Li2003}
\bibinfo{author}{\bibfnamefont{D.}~\bibnamefont{Li}},
  \bibinfo{author}{\bibfnamefont{Y.}~\bibnamefont{Wu}},
  \bibinfo{author}{\bibfnamefont{P.}~\bibnamefont{Kim}},
  \bibinfo{author}{\bibfnamefont{L.}~\bibnamefont{Shi}},
  \bibinfo{author}{\bibfnamefont{P.}~\bibnamefont{Yang}}, \bibnamefont{and}
  \bibinfo{author}{\bibfnamefont{A.}~\bibnamefont{Majumdar}},
  \bibinfo{journal}{Appl. Phys. Lett.} \textbf{\bibinfo{volume}{83}},
  \bibinfo{pages}{2934} (\bibinfo{year}{2003}{\natexlab{a}}).

\bibitem[{\citenamefont{Hochbaum et~al.}(2008)\citenamefont{Hochbaum, Chen,
  Delgado, Liang, Garnett, Najarian, Majumdar, and Yang}}]{Hochbaum2008}
\bibinfo{author}{\bibfnamefont{A.}~\bibnamefont{Hochbaum}},
  \bibinfo{author}{\bibfnamefont{R.}~\bibnamefont{Chen}},
  \bibinfo{author}{\bibfnamefont{R.}~\bibnamefont{Delgado}},
  \bibinfo{author}{\bibfnamefont{W.}~\bibnamefont{Liang}},
  \bibinfo{author}{\bibfnamefont{E.}~\bibnamefont{Garnett}},
  \bibinfo{author}{\bibfnamefont{M.}~\bibnamefont{Najarian}},
  \bibinfo{author}{\bibfnamefont{A.}~\bibnamefont{Majumdar}}, \bibnamefont{and}
  \bibinfo{author}{\bibfnamefont{P.}~\bibnamefont{Yang}},
  \bibinfo{journal}{Nature} \textbf{\bibinfo{volume}{451}},
  \bibinfo{pages}{163} (\bibinfo{year}{2008}).

\bibitem[{\citenamefont{Kim et~al.}(2006)\citenamefont{Kim, Zide, Gossard,
  Klenov, Stemmer, Shakouri, and Majumdar}}]{Kim2006}
\bibinfo{author}{\bibfnamefont{W.}~\bibnamefont{Kim}},
  \bibinfo{author}{\bibfnamefont{J.}~\bibnamefont{Zide}},
  \bibinfo{author}{\bibfnamefont{A.}~\bibnamefont{Gossard}},
  \bibinfo{author}{\bibfnamefont{D.}~\bibnamefont{Klenov}},
  \bibinfo{author}{\bibfnamefont{S.}~\bibnamefont{Stemmer}},
  \bibinfo{author}{\bibfnamefont{A.}~\bibnamefont{Shakouri}}, \bibnamefont{and}
  \bibinfo{author}{\bibfnamefont{A.}~\bibnamefont{Majumdar}},
  \bibinfo{journal}{Phys.\ Rev. Lett.} \textbf{\bibinfo{volume}{96}},
  \bibinfo{pages}{045901} (\bibinfo{year}{2006}).

\bibitem[{\citenamefont{Mingo et~al.}(2009)\citenamefont{Mingo, Hauser,
  Kobayashi, Plissonnier, and Shakouri}}]{Mingo2009}
\bibinfo{author}{\bibfnamefont{N.}~\bibnamefont{Mingo}},
  \bibinfo{author}{\bibfnamefont{D.}~\bibnamefont{Hauser}},
  \bibinfo{author}{\bibfnamefont{N.}~\bibnamefont{Kobayashi}},
  \bibinfo{author}{\bibfnamefont{M.}~\bibnamefont{Plissonnier}},
  \bibnamefont{and} \bibinfo{author}{\bibfnamefont{A.}~\bibnamefont{Shakouri}},
  \bibinfo{journal}{Nano Lett.} \textbf{\bibinfo{volume}{9}},
  \bibinfo{pages}{711} (\bibinfo{year}{2009}).

\bibitem[{\citenamefont{Bera et~al.}(2010)\citenamefont{Bera, Mingo, and
  Volz}}]{Bera2010}
\bibinfo{author}{\bibfnamefont{C.}~\bibnamefont{Bera}},
  \bibinfo{author}{\bibfnamefont{N.}~\bibnamefont{Mingo}}, \bibnamefont{and}
  \bibinfo{author}{\bibfnamefont{S.}~\bibnamefont{Volz}},
  \bibinfo{journal}{Phys.\ Rev. Lett.} \textbf{\bibinfo{volume}{104}},
  \bibinfo{pages}{115502} (\bibinfo{year}{2010}).

\bibitem[{\citenamefont{Li et~al.}(2003{\natexlab{b}})\citenamefont{Li, Wu,
  Fan, Yang, and Majumdar}}]{Li2003a}
\bibinfo{author}{\bibfnamefont{D.}~\bibnamefont{Li}},
  \bibinfo{author}{\bibfnamefont{Y.}~\bibnamefont{Wu}},
  \bibinfo{author}{\bibfnamefont{R.}~\bibnamefont{Fan}},
  \bibinfo{author}{\bibfnamefont{P.}~\bibnamefont{Yang}}, \bibnamefont{and}
  \bibinfo{author}{\bibfnamefont{A.}~\bibnamefont{Majumdar}},
  \bibinfo{journal}{Appl. Phys. Lett.} \textbf{\bibinfo{volume}{83}},
  \bibinfo{pages}{3186} (\bibinfo{year}{2003}{\natexlab{b}}).

\bibitem[{\citenamefont{Kim et~al.}(2010)\citenamefont{Kim, Kim, Choi, and
  Kim}}]{Kim2010}
\bibinfo{author}{\bibfnamefont{H.}~\bibnamefont{Kim}},
  \bibinfo{author}{\bibfnamefont{I.}~\bibnamefont{Kim}},
  \bibinfo{author}{\bibfnamefont{H.}~\bibnamefont{Choi}}, \bibnamefont{and}
  \bibinfo{author}{\bibfnamefont{W.}~\bibnamefont{Kim}},
  \bibinfo{journal}{Appl. Phys. Lett.} \textbf{\bibinfo{volume}{96}},
  \bibinfo{pages}{233106} (\bibinfo{year}{2010}).

\bibitem[{\citenamefont{Mingo}(2003)}]{Mingo2003}
\bibinfo{author}{\bibfnamefont{N.}~\bibnamefont{Mingo}},
  \bibinfo{journal}{Phys.\ Rev.~B} \textbf{\bibinfo{volume}{68}},
  \bibinfo{pages}{1133081} (\bibinfo{year}{2003}).

\bibitem[{\citenamefont{Steele and Rosi}(1958)}]{Steele1958}
\bibinfo{author}{\bibfnamefont{M.}~\bibnamefont{Steele}} \bibnamefont{and}
  \bibinfo{author}{\bibfnamefont{F.}~\bibnamefont{Rosi}}, \bibinfo{journal}{J.
  Appl. Phys.} \textbf{\bibinfo{volume}{29}}, \bibinfo{pages}{1517}
  (\bibinfo{year}{1958}).

\bibitem[{\citenamefont{Abeles et~al.}(1962)\citenamefont{Abeles, Beers, Cody,
  and Dismukes}}]{Abeles1962}
\bibinfo{author}{\bibfnamefont{B.}~\bibnamefont{Abeles}},
  \bibinfo{author}{\bibfnamefont{D.}~\bibnamefont{Beers}},
  \bibinfo{author}{\bibfnamefont{G.}~\bibnamefont{Cody}}, \bibnamefont{and}
  \bibinfo{author}{\bibfnamefont{J.}~\bibnamefont{Dismukes}},
  \bibinfo{journal}{Phys. Rev.} \textbf{\bibinfo{volume}{125}},
  \bibinfo{pages}{44} (\bibinfo{year}{1962}).

\end{thebibliography}
\end{document}